\documentstyle[12pt]{article}
\begin{document}
\renewcommand{\thefootnote}{\fnsymbol{footnote}}
\begin{center}
\begin{large}
\centerline{\bf LEFT and RIGHT HANDEDNESS of FERMIONS and BOSONS}
\end{large}

\baselineskip=16pt
\vspace{0.8cm}
\centerline{\rm NORMA MANKO\v C BOR\v STNIK}
\baselineskip=13pt
\centerline{\it Dept. of Physics, University of
Ljubljana, Jadranska 19, and }
\centerline{\it J. Stefan Institute, Jamova 39,
Ljubljana, 1111, Slovenia }
\vspace{2mm}
\centerline{\rm and}
\centerline{\rm ANAMARIJA BOR\v STNIK}
\baselineskip=13pt
\centerline{\it Department of Physics, University of
Ljubljana, Jadranska 19 }

\vspace{0.9cm}

\abstract{
It is shown, by using Grassmann space to describe the internal
degrees of freedom of fermions and bosons,  that the Weyl like
equation exists not only for massless fermions but also for
massless gauge bosons. The corresponding states have well defined
helicity and handedness. It is also shown that spinors and gauge
bosons  only interact if they are of the same handedness. 
}
\end{center}


\vspace{0.3cm}
\vspace{0.3cm}
\section{  Introduction.}

We discuss in this paper the properties of massless fermions and
gauge bosons, described by the irreducible representations of
the Poincar\' e group. We show that massless spinor and gauge
vector representations exist, induced from those of a little
group, which are eigenstates of the helicity $ h =
\frac{\overrightarrow{M} .
\overrightarrow{p}}{|\overrightarrow{M}.\overrightarrow{p}|} $
and the  
handedness\footnote[1]{In the literature helicity and
handedness are often mixed. Neutrino and antineutrino, for
example, both left handed with $\Gamma = -1, $ have helicity $h
= -1$ and $h = +1$, respectively.} $\Gamma$ operator, with $
M_{i} = \frac{1}{2} 
\epsilon_{ijk} M^{jk}; \;\; 
i,j,k \in \{1,2,3\},( \;\; M^{ab} $ are the generators of the Lorentz
group $ SO(1,3) $, $p_a $ is the four momentum operator and
$\Gamma = \frac{-i}{3!}\; \eta\;\; \epsilon_{abcd} M^{ab}
M^{cd}$, with $\eta = 1 $ for fermions and $\eta = \frac{3}{8} $
for bosons, is 
one of the two invariants of the Lorentz group $SO(1,3)$ with
$\epsilon_{abcd...}$  defined in Eq.(2.7)).
Using Grassmann space to define the internal degrees of
freedom, we show that both, spinors and gauge vectors, fulfil
the operator equation 

$$ \overrightarrow{M} . \overrightarrow{p} = \xi p_0 \Gamma,
\eqno(1.1) $$
\noindent
with $\xi = \frac{1}{2} $ for spinors and $\xi = 1$ for gauge bosons.
The square of Eq.(1.1) leads to the equation
$p^2 = 0$; for spinors this is true on the operator level,  for gauge
vectors only on the representation level.

We further show that the left handed fermions only
interact with the left handed bosons and the right handed
fermions only interact with the right handed bosons.

However, the left-right symmetry is not observed in nature. The photon
and the gluon have 
well defined internal parity rather than handedness. They  can
accordingly be 
described by the sum of states with well defined
handedness, which still  have well defined helicity $h$ but
they no longer fulfil the Weyl like equation (1.1).
Massless fields have, no matter whether
their spin is $\frac{1}{2} $ or $1$, two dimensional
representations of the Lorentz group\cite{wig,sch,fonda}. 
Because of that the gauge  bosons (which are the three rather
than the four vectors\cite{bd,gour,wein,fonda,han,man1,man2})
manifest the gauge invariance, generated by the little group
generators\cite{gour,han,fonda,mirman}.

We  discuss in this paper those properties of  massless
fermions and gauge bosons, which are connected with only spin
degrees of freedom and are determined by the generators of the Lorentz
transformations,  the homogeneous and the 
discrete, of the group $SO(1,3)$ embedded into the group
$SO(1,4)$ and are part of the Poincar\' e group. 

To understand why the left-right symmetry is the broken symmetry
in nature, one has to study not only spins but also charges\footnote[2]  
{The standard electroweak model postulates in order to be in
agreement with the experimental data that only left handed
fermions  carry the weak charge, while the right handed
fermions do not. Accordingly right handed fermions do not
interact with the weak bosons. }.
The authors of refs.\cite{pati,moha} 
restore the left-right symmetry in a phenomenological way, they
label groups, introduced to describe charges,  by the
handedness: Their two weak charge $SU(2)$ groups carry 
representations of the left and  right handedness,
respectively, while the hypercharge group 
$U(1)$ and the $SU(3)$ colour charge group carry the
representations of the left plus right symmetry. We shall
comment on this point in the last section. We believe that the
break of the left-right symmetry can only be understood within
the unified theory of spins and charges.
For this purpose the approach was introduced\cite{man2,man3},
which  describes all the internal degrees of freedom of
fermions and bosons in an unique way. The approach assumes that
space has d ordinary commuting and d Grassmann anticommuting
coordinates, with $ d\ge 15$. 
Since  Grassmann space enables to define two kinds of generators
of the Lorentz transformations, one of spinorial
the other of vectorial character,  two different  kinds of
representations exist, appropriate to describe spins and
charges of fermions and gauge bosons,
respectively\cite{man1,man2,man3}.  
For $d = 15$, contain the representations, if
determined by the operators forming the subgroups $SO(1,4),
SU(3), SU(2), U(1)$ of the group $SO(1,14)$, all the vectors
needed to describe spins and charges of all known fermions and
gauge bosons, as well as the scalars\cite{man1}.
Spins and charges are in the proposed theory unified, separately
for spinors and separately for vectors.

Although many of the properties presented in this paper can
well be understood in the usual group theoretical approach, the
Grassmann space approach, with generators of groups represented
by the differential operators in Grassmann space\cite{man2,man1,man3},
offers the transparent way of understanding the behaviour of 
particles with respect to the internal degrees of
freedom.

\section{  Coordinate Grassmann space, linear operators,
Lorentz group SO(1,3) and representations.}

In this section we present the representations of the Lorentz
group for spinors and vectors as polynomials in Grassmann space.
We find spinors to form four times the two Weyl spinors, one of the
left ($\Gamma = -1 \;\;)$ (the $(\frac{1}{2}, 0)$ representation), the
other of the right ($\Gamma = 1)\;\;$ ( the $(0, \frac{1}{2})$
representation) handedness. The matrix representations for the
generators of the 
Lorentz group and the invariants of the group are the same for
all the four cases (Eq.2.11a). We find vectors to form a true
scalar $( \Gamma = 0 ) $ and a pseudo scalar $ (\Gamma = 0 )$
representation with zero handedness, the three 
left handed vectors $\;\;(\Gamma = -1)\;\;)$ ( the $(1, 0)$
representation), the three right handed vectors $\;\;(\Gamma =
1$) ( the $(0, 1)$ representation), and the two four vectors of zero
handedness $\;\;(\Gamma = 0)\;\;$ (the $(\frac{1}{2},
\frac{1}{2})$ representations).  

We first briefly repeat a few definitions concerning
d-dimensional Grassmann space,  linear Grassmann space 
spanned over the space of anticommuting coordinates, linear
operators defined in 
this space and the Lie algebra of generators of the Lorentz
transformations \cite{man1,man2,man3,ber}. We solve the
eigenvalue problem and present the representations.

\subsection{Coordinate space with Grassmann character.}

We define a d-dimensional Grassmann space of real anticommuting
coordinates $ \{ \theta ^a \} $, $ a=0,1,2,3,5,6,...,d,$
satisfying the anticommutation relations
$ \theta^a \theta^b + \theta^b \theta^a := \{ \theta^a, \theta^b
\} = 0,$
called the Grassmann algebra \cite{man1,man3,ber}. The
metric tensor $ \eta 
_{ab}$ $ = diag (1,$ $ -1, -1,$ $ -1,..., -1) $ lowers the
indices of a 
vector $\{ \theta^a \} = \{ \theta^0, \theta^1,..., \theta^d \},
\theta_a = \eta_{ab} \theta^b$. Linear transformation actions on
vectors $ (\alpha \theta^a + \beta  x^a),\;\; $
$ (\alpha \acute{\theta}^a + \beta \acute{x}^a ) = L^a{ }_b
(\alpha \theta^b + \beta x^b ), $
which leave forms
$ ( \alpha \theta^a + \beta x^a ) ( \alpha \theta ^b + \beta
x^b ) \eta_{ab} $
invariant, are called the Lorentz transformations. Here $
(\alpha \theta^a + \beta x^a ) $ is a  vector of d
anticommuting components  and d commuting $ (x^ax^b -
x^bx^a = 0) $ components, and $ \alpha $ and $ \beta$ are two
complex numbers. The requirement that forms $ ( \alpha \theta^a
+ \beta x^a ) ( \alpha \theta ^b + \beta 
x^b ) \eta_{ab} $ are scalars with
respect to the above linear transformations, leads to the
equations 
$ L^a{ }_c L^b{ }_d \eta_{ab} = \eta_{cd}.$

\subsection{Linear vector space.}

A linear space spanned over a Grassmann coordinate space of d
coordinates has the dimension $ 2^d$. If monomials $
\theta^{\alpha_1} \theta^{\alpha_2}....\theta^{\alpha_n} $
are taken as a set of basic vectors with $\alpha_i \neq
\alpha_j,$ half of the vectors have an even (those with an even n)
and half of the vectors have an odd (those with an odd n)
Grassmann character. Any vector in this space may be represented
as a linear superposition of monomials 

$$ f(\theta) = \alpha_0 + \sum_{i=1}^{d}  \alpha _{a_1a_2 ..a_i}
\theta^{a_1} \theta^{a_2}....\theta^{a_i},\;\; a_k< a_{k+1},
\eqno(2.1)$$ 
\noindent
where constants $\alpha_0, \alpha_{a_1a_2..a_i}$ are complex
numbers. 

\subsection{Linear operators.}

In  Grassmann space the left derivatives have to be
distinguished  from the right derivatives, due to the 
anticommuting nature of the coordinates \cite{man1,man3,ber}. We
shall make use 
of left derivatives $ {\overrightarrow {{\partial}^{\theta}}}{
}_a:= \frac{\overrightarrow {\partial}}{\partial
\theta^a},\;\;\; {\overrightarrow {{\partial}^{\theta}}}{ }^{a}:=
{\eta ^{ab}} \overrightarrow {{\partial}^{\theta}}{ }_b \; $, on
vectors of the 
linear space of monomials $ f(\theta)$,  defined as follows:
$ {\overrightarrow{{\partial}^{\theta}}}{ }_a\; \theta^b f(\theta)
= \delta^b{ }_a f(\theta) - \theta^b  
{\overrightarrow{{\partial}^{\theta}}}{ }_a\; f(\theta).$

We define the  linear operators \cite{man1,man3}

$$ p^{\theta} { }^a := i {\overrightarrow{{\partial}^{\theta}}}{
}^a , \;\;
 \tilde{a} ^a := -i(p^{\theta a} + i \theta^a) ,\;\;
\tilde{\tilde{a}}{}^a := (p^{\theta a} - i \theta^a). \eqno
(2.2) $$

According to the inner product defined in Eq.(2.8),
the operators $  \tilde{a} ^a $ and $ \tilde{\tilde{a}}{}^a $
are either hermitian or antihermitian operators
$\;\; \theta^{a+} = i\eta^{aa} p^{\theta a},\;\; p^{\theta a +}
= -i\eta^{aa} 
p^{\theta a},\;\; \tilde{a}^{a +} = \eta^{aa} \tilde{a}^{a},\;\;
\tilde{\tilde{a}}^{a +} = \eta^{aa} \tilde{\tilde{a}}^{a}.  $

We define the generalized commutation relations
(which follow from the corresponding Poisson brackets
\cite{man1,man2,man3}): 

$$ \{ A,B \} := AB - (-1) ^{n_{AB}} BA, \eqno(2.3) $$
\noindent
fulfilling the equation  
$ \{A,B\}  = (-1)^{n_{AB}+1} \{B,A\}.$  
Here $n_{AB}$ is defined as follows

$$ n_{AB} = \left\{ \begin{array} {ll} +1, & if\; A\; and\; B \;
have\;Grassmann\; odd\; character\\
0, & otherwise \end{array} \right\}. $$

We find
$$ \{p^{\theta a}, p^{\theta b} \} = 0 = \{ \theta^{a},
\theta^{b}\},\;\; \{p^{\theta a}, \theta^{b}\} = i \eta^{ab},\;\;
\{\tilde{a}^{a}, \tilde{a}^{b} \} = 2 \eta^{ab} 
= \{\tilde{\tilde{a}}{ }^{a}, \tilde{\tilde{a}}{ }^{b} \},\;\; \{
\tilde{a}^{a}, \tilde{\tilde{a}}{ }^{b} \} = 0. \eqno(2.4)$$ 

\noindent
We see that $\theta ^a $ and $ p^{\theta a} $ form a Grassmann
odd Heisenberg algebra, while $ \tilde a^a $ and $
\tilde{\tilde{a}}{ }^a $ form the Clifford algebra.
In this paper we shall make use of only $\tilde{a}^a$. We shall
put $\tilde{\tilde{a}}^a $
equal zero. We shall comment on this point in Sect. 3.

\subsection{Lie algebra of Lorentz group.}

We define two kinds of operators \cite{man1,man2}. The first ones are
binomials of operators forming the Grassmann odd Heisenberg
algebra 

$$ S^{ab} : = ( \theta^a p^{\theta
b} - \theta ^b p^{\theta a} ).                     \eqno  (2.5a)$$
\noindent
The second ones are binomials of operators forming the Clifford
algebra 

$$ \tilde S ^{ab}: =  \frac{i}{4} [\tilde a ^a , \tilde a ^ b
],  \eqno(2.5b)$$
\noindent
with $ [A, B]:= AB - BA.$  

Either the operators $ S^{ab} $ or $ \tilde S ^{ab}$  
fulfil the Lie algebra of the Lorentz group $ SO(1,d-1)
$ in  d-dimensional Grassmann space:
$$ \{ M^{ab}, M^{cd} \} = i ( M^{ad} \eta^{bc} + M^{bc}
\eta^{ad} - M^{ac} \eta^{bd} - M^{bd} \eta^{ac} ),\eqno(2.6)$$

\noindent
with $ M^{ab} $ equal either to $ S ^{ab} $ or to $\tilde S
^{ab} $
and $ M^{ab} = -M^{ba}. $   

By solving the eigenvalue problem (see below) we find that
operators $ \tilde S ^{ab} $ define   
the spinorial representations of the Lorentz group, while
$S^{ab} $ define the  vectorial
representations of the Lorentz group $ SO(1,d-1) $.

Group elements are in any of the two cases defined by:
$ {\cal U}(\omega) = e^{-\frac{i}{2} \omega_{ab} M^{ab}},$
where $ \omega_{ab} $ are the parameters of the group.

It can be proved for any d that
$ M^2 $ is the invariant of the Lorentz group 
$ \{ M^2, M^{cd} \} = 0,\;\; M^2 =  \frac{1}{2} M^{ab} M_{ab},$ 
and that for d=2n we can find an additional invariant $ \Gamma $

$$  \Gamma = \eta \frac{i(-2i)^{n}
}{(2n)!} 
\epsilon_{a_1a_2...a_{2n}} M^{a_1a_2}
....M^{a_{2n-1}a_{2n}},\;\;\{ \Gamma, M^{cd} \} = 0, 
 \eqno(2.7) $$

\noindent
where $\epsilon _{a_1a_2...a_{2n}} $ is the totally antisymmetric
tensor with $ 2n $ indices and with $ \epsilon _{ 1 2 3 ...2n }
= 1$ and $\eta$ is a real constant. We choose $\eta = 1$ for
spinors and $\eta = \frac{3}{8}$ for vectors.
This means that $ M^2  $ 
and $ \Gamma $ are for $ d = 4$ the two invariants or Casimir
operators of the group $ SO(1,3). $

While the invariant $ M^2 $ is trivial in the
case when $ M^{ab} $ has spinorial character, since  
$ (\tilde S^{ab})^2 = \frac{1}{4} \eta^{aa} \eta^{bb}$ and
$ M^2 $ is equal
to a number $ \frac{1}{2} \tilde
S^{ab} \tilde S _{ab} = \frac{3}{2}$ , it is a nontrivial
differential operator in  Grassmann space if $ M^{ab}$
have  vectorial character $(M^{ab} = S^{ab})$. The invariant of
Eq.(2.7) is always a nontrivial operator.  

\subsection{Integrals on Grassmann space.}

We assume that differentials of Grassmann coordinates $ d\theta^a
$ fulfil the Grassmann anticommuting relations \cite{man1,man3,ber}
$ \{ d\theta^a, d\theta^b \} = 0$
and we introduce a single integral over the whole interval of
$d\theta^a: \;\; $ 
$ \int d\theta^a = 0, \;\; \int d\theta^a \theta^a = 1,\;\; a =
0,1,2,3,5,..,d,$
and the multiple integral over d coordinates
$ \int d^d \theta \theta^0 \theta^1 \theta^2 \theta^3
\theta^4...\theta^d = 1,$ with 
$ d^d \theta: = d\theta^d...d\theta^3 d\theta^2 d\theta^1
d\theta^0 $ in the standard way.

We define \cite{man1,man3,ber} the inner product of two vectors $
\langle\varphi|\theta\rangle $ and $ \langle\theta|\chi\rangle $, with $
\langle\varphi|\theta\rangle = \langle\theta|\varphi\rangle^* $ as follows:

$$ \langle\varphi|\chi\rangle = \int d^d\theta ( \omega\langle\varphi|\theta\rangle)
\langle\theta| \chi\rangle, \eqno(2.8) $$

\noindent
with the weight function 
$\omega = \prod_{k=0,1,2,3,..,d}
(\frac{\partial}{\partial \theta^k}  + \theta^k ),$ 
which operates on the first function $ \langle\varphi|\theta\rangle $ only, 
and we define
$ (\alpha^{a_1 a_2...a_k} \theta^{a_1}
\theta^{a_2}...\theta^{a_k})^{+} =
(\theta^{a_k}).....(\theta^{a_2}) 
(\theta^{a_1}) (\alpha^{a_1 a_2...a_k})^{*}.$

\noindent
According to the above definition of the inner product 
the generators of the Lorentz 
transformations (Eqs.(2.5b)) are self adjoint ( if $
a \neq 0 $ and  $ b \neq 0 $ ) or antiself adjoint ( if $ a =
0 $ or $ b = 0 $ )  operators.

The volume element $ d^d\theta $ and the weight function $
\omega$ are invariants with respect to the Lorentz
transformations,  both are scalar densities of weight $- 1$.

According to Eqs.(2.2) and (2.5) we find

$$ S^{ab} = i ( \theta^a \frac{\partial}{\partial \theta_b}
- \theta^b \frac{\partial}{\partial \theta_a} ),\;\; \tilde a^a
= (\frac{\partial}{\partial \theta_a} + 
\theta^a ),\;\; \tilde S ^{ab} = \frac{i}{2}( \frac{\partial}{\partial
\theta_a} + 
\theta^a ) ( \frac{\partial}{\partial \theta_b} +
\theta^b ),\;if\;a \neq b.\eqno(2.9)  $$

\subsection{Eigenvalue problem.}

To find  eigenvectors of any operator $A$, we
solve the eigenvalue problem 

$$ \langle\theta|\tilde A_i|\tilde{\varphi}\rangle = \tilde{\alpha}_i
\langle\theta|\tilde{\varphi}\rangle ,\;\;
\langle\theta|A_i|\varphi\rangle = \alpha_i \langle\theta|\varphi\rangle,\;\;i = \{1,p\}
,\eqno(2.10)$$

\noindent
where $ \tilde{A}_i $ and $ A_i $ stand for $p$ commuting
operators 
of spinorial and vectorial character, respectively and
$\tilde{\alpha}_i\;$  and $\alpha
_i$ for the corresponding eigenvalues.

To solve  equations (2.10) we express the operators in the
coordinate representation  and write the eigenvectors
as polynomials of $\theta^a$. We  orthonormalize the vectors
according to the inner product, defined in Eq.(2.8), 
$ \langle{ }^{\alpha} \tilde{\varphi}_i |{ }^{\beta} \tilde{\varphi}_j\rangle =
\delta^{\alpha \beta} \delta_{ij},\;\;\; \langle{ }^{\alpha}
{\varphi}_i |{ }^{\beta} 
{\varphi}_j\rangle = \delta^{\alpha \beta} \delta_{ij},$
where index $\alpha $ distinguishes between vectors of different
irreducible representations and index j between vectors of the
same irreducible representation. This determines the
orthonormalization condition for spinorial and vectorial
representations, respectively.

\subsection{Representations of spinors and vectors of $SO(1,3)$
embedded into $SO(1,4)$.}

We solve the eigenvalue problem for the invariants $M^2$ and
$\Gamma $ as well as for
the commuting operators $M^{12}$ and $M^{03}$ of the Lorentz
group $SO(1,3)$ for either
spinorial or vectorial operators (Eqs.(2.9)). We take $d = 5$ for
the reason that 
the operator of an even Grassmann character $\tilde{\gamma} =
-2i \tilde{S}^{5m} = \tilde{a}^5 \tilde{a}^m, m \in \{0,1,2,3 \}
$ can be defined, with all the properties of the Dirac
$\gamma^m$ operators\footnote[3]{ Having an even Grassmann
character $\tilde{\gamma}^a$ transform spinors into spinors,
otherwise they would change the Grassmann character of fields,
causing the supersymmetric transformations\cite{man1}.}. In $d =
5$ 
dimensional Grassmann space 
there are 16 vectors of an odd and 16 vectors of an even
Grassmann character. We find within the odd vectors the
irreducible representations of the Lorentz group $SO(1,3)$ for
the spinorial operators and within the even vectors the
irreducible representations of the Lorentz group $SO(1,3)$ for
the vectorial operators\footnote[4]{ The reason for such a
choice is that in the canonical quantization of fields the
Grassmann odd vectors
quantize to fermions, the Grassmann even  to bosons.}.

In the spinorial case we find four times two spinors, one left
$(\Gamma = -1)$ and one 
right $(\Gamma = 1)$ handed, each two spinor coupled to a bispinor
by the operator $\tilde{\gamma}^m$. The four different two
spinor representations distinguish among themselves with respect
to the discrete transformations of the Lorentz group $SO(1,3)$:
$\overrightarrow{\theta} \longrightarrow - \overrightarrow{\theta}$,
$\theta^0 \longrightarrow - \theta^0$ and the complex
conjugation. We present these representations in Table I. We
introduce the notation: $ \tilde{S}_i =
\frac{1}{2} \epsilon_{ijk} \tilde{S}^{jk},\;\; \tilde{K}_i =
\tilde{S}^{0i}$. In Eq.(2.11a) we present the matrix
representation for the generators of the Lorentz transformations
of the group $SO(1,3)$, for the invariant $\tilde{\Gamma}$ and
for $\tilde{\gamma}^0$ for any of the four vectors.
Starting from any  of the sixteen vectors, all the others  can be
obtained by applying on the starting one
either the generators $\tilde{S}^{mn}$, or  $\tilde{\gamma}^m$
or  the generators of the discrete symmetries: space
inversion, time inversion or complex conjugation.

\vspace{3mm}

\begin{center}
\begin{tabular}{|c|c||c||r|r|r|c|}
\hline
$ \alpha$& $i$ & 
$\langle\theta|{ }^{\alpha} \tilde{\varphi}_{i}\rangle$ &  $\tilde{S}^{3}$ & 
$\tilde{K}^{3}$ & $ \tilde{\Gamma} $&
$\tilde{\cal S}^2$  \\
\hline
\hline 
$1$ & $1$ & $ -\frac{1}{2}
( \theta^{1} + i \theta^{2})(\theta^{0} + \theta^{3})\theta^{5}$
& 
$ \frac{1}{2}$ & $ -\frac{i}{2}$ &$ -1$ & $\frac{3}{2}$  \\
\hline
$1$ & $2$ & $ \frac{1}{2}(1 - 
i \theta^{1}\theta^{2})(1 + \theta^{0}\theta^{3})\theta^{5}$ &
$ -\frac{1}{2}$ & $ \frac{i}{2}$ &$ -1$ & $\frac{3}{2}$ \\
\hline
$2$ & $1$ & $ \frac{1}{2}
(\theta^{1} + i \theta^{2})(1 + \theta^{0}\theta^{3})$ & 
$ \frac{1}{2}$ & $ \frac{i}{2}$ &$ 1$  
&$ \frac{3}{2} $  \\
\hline
$2$ & $2$ & $ -\frac{1}{2}(1 - 
i \theta^{1}\theta^{2})(\theta^{0} + \theta^{3})$ &
$ -\frac{1}{2}$ & $ -\frac{i}{2}$ &$ 1$ 
&$ \frac{3}{2} $  \\
\hline\hline 
$3$ & $1$ & $ -\frac{1}{2}
( \theta^{1} + i \theta^{2})(1 - \theta^{0}\theta^{3})$ & 
$ \frac{1}{2}$ & $ -\frac{i}{2}$ &$ -1$ & $\frac{3}{2}$  \\
\hline
$3$ & $2$ & $ \frac{1}{2}(1 - 
i \theta^{1}\theta^{2})(\theta^{0} - \theta^{3})$ &
$ -\frac{1}{2}$ & $ \frac{i}{2}$ &$ -1$ & $ \frac{3}{2}$ \\
\hline
$4$ & $1$ & $ \frac{1}{2}
( \theta^{1} + i \theta^{2})(\theta^{0} - \theta^{3})\theta^{5}$
& 
$ \frac{1}{2}$ & $ \frac{i}{2}$ &$ 1$ 
&$ \frac{3}{2} $   \\
\hline
$4$ & $2$ & $ \frac{1}{2}(1 - 
i \theta^{1}\theta^{2})(1 - \theta^{0}\theta^{3})\theta^{5}$ &
$ -\frac{1}{2}$ & $ -\frac{i}{2}$ &$ 1$ 
&$ \frac{3}{2} $\\
\hline\hline
$5$ & $1$ & $ -\frac{1}{2}(1 +
i \theta^{1}\theta^{2})(\theta^{0} + \theta^{3})$ &
$ \frac{1}{2}$ & $ -\frac{i}{2}$ &$ -1$ & $ \frac{3}{2}$  \\
\hline
$5$ & $2$ & $ \frac{1}{2}(\theta^{1} - i \theta^{2})(1 +
\theta^{0}\theta^{3})$ & 
$ -\frac{1}{2}$ & $ \frac{i}{2}$ &$ -1$ & $ \frac{3}{2}$ \\
\hline
$6$ & $1$ & $ -\frac{1}{2}(1 + i \theta^{1}\theta^{2})(1 +
\theta^{0}\theta^{3})\theta^{5}$ & 
$ \frac{1}{2}$ & $ \frac{i}{2}$ &$ 1$ 
&$ \frac{3}{2} $ \\
\hline
$6$ & $2$ & $ \frac{1}{2}(\theta^{1} - i \theta^{2})(\theta^{0}
+ \theta^{3})\theta^{5}$ & 
$ -\frac{1}{2}$ & $ -\frac{i}{2}$ &$ 1$ 
&$ \frac{3}{2} $ \\
\hline
\hline
$7$ & $1$ & $ -\frac{1}{2}(1 +
i \theta^{1}\theta^{2})(1 - \theta^{0}\theta^{3})\theta^{5}$ &
$ \frac{1}{2}$ & $ -\frac{i}{2}$ &$ -1$ & $ \frac{3}{2}$  \\
\hline
$7$ & $2$ & $ \frac{1}{2}(\theta^{1} - i \theta^{2})(\theta^{0}
- \theta^{3})\theta^{5}$ & 
$ -\frac{1}{2}$ & $ \frac{i}{2}$ &$ -1$ & $ \frac{3}{2}$ \\
\hline
$8$ & $1$ & $ -\frac{1}{2}(1 + i \theta^{1}
\theta^{2})(\theta^{0} - \theta^{3})$ & 
$ \frac{1}{2}$ & $ \frac{i}{2}$ &$ 1$
& $ \frac{3}{2} $ \\
\hline
$8$ & $2$ & $ -\frac{1}{2}(\theta^{1} - i \theta^{2})(1 -
\theta^{0}\theta^{3})$ & 
$ -\frac{1}{2}$ & $ -\frac{i}{2}$ &$ 1$ 
&$ \frac{3}{2} $ \\
\hline
\end{tabular}
\end{center}

\vspace{3mm}

Table I. Irreducible representations of the Lorentz group
$SO(1,3)$ embedded into the group $SO(1,4)$. The polynomials
demonstrate the $SU(2) \times SU(2)$ structure of the Lorentz group
$SO(1,3)$ and represent four times the left and the right handed
Weyl spinor\cite{man1}. We use the notation $\tilde{\cal S}^2 =
\frac{1}{2} \tilde{S}^{ab} \tilde{S}_{ab}. $

\vspace{3mm}

$$
\tilde{\Gamma}_{\alpha} = \left( 
\begin{array}{rr} -I&0\\ 0& I \end{array} \right),\;\;
\overrightarrow{\tilde{S}}_{\alpha} = \left( 
\begin{array}{rr} \overrightarrow{\tilde{s}}&0\\ 
0& \overrightarrow{\tilde{s}} \end{array} \right),\;\;
\overrightarrow{\tilde{K}}_{\alpha} = \left( 
\begin{array}{rr} -i\overrightarrow{\tilde{s}}&0\\ 
0& i\overrightarrow{\tilde{s}} \end{array} \right),$$
$$
\tilde{\gamma}^0_{\alpha} = \left( 
\begin{array}{rr} 0&I\\ I& 0 \end{array} \right),\;\;
\overrightarrow{\tilde{s}} = \frac{1}{2}
\overrightarrow{\sigma}, \;\;
I = \left( 
\begin{array}{rr} 1&0\\ 0& 1 \end{array} \right),$$
$$ {\rm here }\;\; \overrightarrow{\sigma} \;\; {\rm are\;\;
Pauli \;\; matrices.\;\; One\;\; finds \;\;}
\overrightarrow{\tilde{K}}_{\alpha} = i \tilde{\Gamma}_{\alpha}
\overrightarrow{\tilde{S}}_{\alpha}.
\eqno(2.11a)$$

\vspace{3mm}

In the vectorial case we find a true and a pseudo scalar of zero
handedness, one three vector of the left handedness ($\Gamma =
-1$), one three vector of the right handedness ($\Gamma = 1$)
and two four vectors of zero handedness. We present the results
in Table II, using the equivalent notation as in the spinorial
case. In Eq.(2.11b) we present the matrix representations of
the operators for the two three vectors of different
handedness. While the matrices are in the spinorial 
case of the dimension four times four\footnote[5]{ The matrices
are in the spinorial case of the dimension sixteen times
sixteen, 
but having the block diagonal structure of dimension four times
four, they are the direct sum of four times the four square matrices.
In the vectorial case the corresponding sixteen times sixteen
matrices are  the direct sum of one six times six matrix, two one
dimensional matrices and one eight times eight 
matrix.}, are in the vectorial case of the dimension six times
six\footnote[6]{ One can find also in the case of the
two three vectors the matrices $\gamma^0$, which transform the left
handed vectors into the right handed ones or opposite, but we
don't find the operator realization of such matrices.}.

\vspace{3mm}

\begin{center}
\begin{tabular}{|c|c||c||c|r|r|r|}
\hline
$ \alpha$& $i$ & 
$\langle\theta|{ }^{\alpha}\varphi_{i}\rangle$ &    ${S}^{3}$ & 
${K}^{3}$ & $  \Gamma $  & ${\cal S}^2$\\
\hline
\hline 
$1$ & $1$ & $ \frac{1}{2}(\theta^{0} + \theta^{3})(\theta^{1} +
i \theta^{2})$ & 
 $ 1$ & $ -i$ & $-1$ & $ 4 $\\
\hline
$1$ & $2$ & $- \frac{1}{{\sqrt 2}}(i\theta^{1}\theta^{2} -
\theta^{0} \theta^{3})$ & 
 $ 0$ & $ 0$ & $-1$  & $ 4$\\
\hline
$1$ & $3$ & $ -\frac{1}{2}(\theta^{0} - \theta^{3})(\theta^{1} -
i \theta^{2})$ & 
 $ -1$ & $ i$ & $-1$ & $4$\\
\hline
\hline
$2$ & $1$ & $ \frac{1}{2}(\theta^{0} - \theta^{3})(\theta^{1} +
i \theta^{2})$ & 
 $ 1$ & $ i$ & $1$ &$4$\\
\hline
$2$ & $2$ & $- \frac{1}{{\sqrt 2}}(i\theta^{1}\theta^{2} +
\theta^{0} \theta^{3})$ & 
 $ 0$ & $ 0$ & $1$ &$4$ \\
\hline
$2$ & $3$ & $ -\frac{1}{2}(\theta^{0} + \theta^{3})(\theta^{1} -
i \theta^{2})$ & 
 $ -1$ & $ -i$ & $1$ &$4$ \\
\hline
\hline
$3$ & $1$ & $ 1 $ &  $ 0$ & $0$ & $0$ &$0$ \\
\hline
$4$ & $1$ & $ i \theta^{0}\theta^{1}\theta^{2}\theta^{3} $ & $
0$ & $ 0$ & $ 0$ & $0$  \\
\hline
\hline
$5$ & $1$ & $ \frac{1}{2}(1 + \theta^{0}\theta^{3})(\theta^{1} -
i \theta^{2}) \theta^5$ & 
 $ -1$ & $ 0$ & $0$ &$3$ \\
\hline
$5$ & $2$ & $ \frac{1}{2}(\theta^{0} - \theta^{3})(1 -
i\theta^{1} \theta^{2}) \theta^5$ & 
 $ 0$ & $ i$ & $0$ &$3$ \\
\hline
$5$ & $3$ & $ \frac{1}{2}(1 - \theta^{0} \theta^{3})(\theta^{1}
+ i \theta^{2}) \theta^5$ & 
 $ 1$ & $ 0$ & $0$ &$3$\\
\hline
$5$ & $4$ & $ \frac{1}{2}(\theta^{0} + \theta^{3})(1 +
i\theta^{1} \theta^{2}) \theta^5$ & 
$ 0$ & $ -i$ & $0$ &$3$ \\
\hline
\hline
$6$ & $1$ & $ \frac{1}{2}(1 - \theta^{0}\theta^{3})(\theta^{1} -
i \theta^{2}) \theta^5$ & 
 $ -1$ & $ 0$ & $0$ &$3$ \\
\hline
$6$ & $2$ & $ \frac{1}{2}(\theta^{0} - \theta^{3})(1 +
i\theta^{1} \theta^{2}) \theta^5$ & 
 $ 0$ & $ i$ & $0$ &$3$\\
\hline
$6$ & $3$ & $ \frac{1}{2}(1 + \theta^{0} \theta^{3})(\theta^{1}
+ i \theta^{2}) \theta^5$ & 
 $ 1$ & $ 0$ & $0$ &$3$ \\
\hline
$6$ & $4$ & $ \frac{1}{2}(\theta^{0} + \theta^{3})(1 -
i\theta^{1} \theta^{2}) \theta^5$ & 
 $ 0$ & $ -i$ & $0$ &$3$\\
\hline
\end{tabular}
\end{center}

\vspace{3mm}

Table II. Irreducible representations of the Lorentz group
$SO(1,3)$ embedded into the group $SO(1,4)$. The polynomials in
Grassmann space represent the two scalars, the two three vectors
and the two four vectors\cite{man1}. We use the notation ${\cal
S} = \frac{1}{2} S^{ab} S_{ab} $.

\vspace{3mm}

$$
\Gamma = \left( 
\begin{array}{rr} -I&0\\ 0& I \end{array} \right),\;\;
\overrightarrow{S} = \left( 
\begin{array}{cc} \overrightarrow{s}&0\\ 
0& \overrightarrow{s} \end{array} \right),\;\;
\overrightarrow{K} = \left( 
\begin{array}{cc} -i\overrightarrow{s}&0\\ 
0& i\overrightarrow{s} \end{array} \right),$$
$$
s_1 = \frac{1}{\sqrt{2}}\left( 
\begin{array}{ccc} 0 & 1 & 0\\ 1 & 0 & 1 \\ 0 & 1 & 0
\end{array} \right),\;\;
s_2 = \frac{1}{\sqrt{2}}\left( 
\begin{array}{rrr} 0 & -i & 0\\ i & 0 & -i \\ 0 & i & 0
\end{array} \right),\;\;
s_3 = \left( 
\begin{array}{ccc} 1 & 0 & 0\\ 0 & 0 & 0 \\ 0 & 0 & -1
\end{array} \right).
\eqno(2.11b)$$
$$
{\rm It\;\; follows\;\; again\;\; that} \overrightarrow{K} = i
\Gamma \overrightarrow{S}. 
$$

\vspace{3mm}

\section{ Particles in ordinary and Grassmann space.}

A particle which lives in a space of d ordinary commuting 
and d Grassmann anticommuting coordinates $X^a:=
\{x^a,\theta^a\} $ moves along the supergeodesics\cite{man3}.
Choosing the appropriate action\cite{man3,ikem} to describe the
dynamics of such a particle, one  derives the two 
Euler-Lagrange equations\cite{man3}:
$ \frac{dp^a}{d \tau} = 0,\;\;\; \frac{dp^{\theta a}}{d \tau} =
 \frac{i}{2}${\it m}$p^a, $
and the three constraints:

$$ \chi^1: = p_a \tilde{a}^a = 0,\;\; \chi^2: = p^a p_a = 0, \;\;
\chi^3{ }_a: = \tilde{\tilde{a}}{ }^a:= p^{\theta a} - i \theta^a
= 0, \;\;{\rm with} 
\tilde{a}^a:= -i p^{\theta a} + \theta^a,
\eqno(3.1)$$

\noindent
where $p^a$ is the conjugate momentum to the coordinate $x^a$,
$p^{\theta a}$ the conjugate momentum to the coordinate
$\theta^a$ and {\it m} is the parameter of the action. One also finds
the generators of the Lorentz transformations:

$$ M^{ a b} = L^{a b} + \tilde{S}^{a b} \;,\; L^{a b} = x^a p^b
- x^b p^a 
\;,\; \tilde{S}^{a b} = \frac{i}{4} (\tilde{a}^{a} \tilde{a}^{b}
- \tilde{a}^{b} \tilde{a}^{a}),
\eqno (3.2) $$

\noindent 
demonstrating that parameters of the Lorentz transformations are the
same in both spaces and that coordinates in Grassmann space are
proportional to their conjugate momenta.

In the canonical quantization of coordinates\cite{man3}, the
momenta $p^a$ and $p^{\theta a} $ and accordingly quantities $
\tilde{a}^a $ and $\tilde{\tilde{a}}{ }^a$ become operators,
fulfilling Eqs. (2.2,2.4) of Sect.2.3.. The two constraints of
Eq.(3.1) 
lead to the Dirac and Klein-Gordon equations, provided that 
coordinates $\theta^a $ and their conjugate momenta $p^{\theta
a}$, which appear in the equations of motion, are replaced by $
\frac{1}{2} \tilde{a}^a$ and $\frac{1}{2i}
\tilde{a}^a$, respectively, while 
$\tilde{\tilde{a}}{ }^a $ are put equal to zero, and that the
generator of the Lorentz transformations $ -2i \tilde{S}^{5
m},\; m = 0,1,2,3,\;\; $ ( having an even Grassmann character,
the operator
doesn't change the Grassmann character of spinors ) is
recognised as the Dirac $\gamma^m$ 
matrices:  $\tilde{\gamma}{ }^m p_{m} |\tilde{\varphi}\rangle = 0 \;,\;
m=0,1,2,3. $
The third constraint of Eq.(3.1) has to be taken  into account 
in the  expectation value form 
$ \langle \tilde{\varphi} | \tilde{\tilde a}{ }^m | \tilde{\varphi} \rangle = 0.
$

In the presence of gauge fields the kinetic momentum $p_m$ in
the Dirac equation has
to be replaced by the canonical momentum: $p_m	 \longrightarrow
p_m{ }_0 = p_m - A_m$, where $A^m$ stays for gauge
fields\cite{man1,man2,man3}.

\section{ Poincar\' e group and Weyl-like equations for
$\;\;\;\;\;\;$  fermions and bosons.}

We are studying  symmetries, connected with the transformation
properties of ordinary and Grassmann coordinates. Besides the
transformations of the Lorentz group $SO(1,3)$ in ordinary and
Grassmann 
space, we allow translations in only ordinary 
space\footnote[7]{ Translations in Grassmann space
cause\cite{man1}  supersymmetric transformations, which shall
be not studied here.}. For the homogeneous transformations we
have: $x^{'a} = L^a{ }_b x^b + b^{b},\;\; 
\theta^{'a} = L^a{ }_b \theta^b,$ where $ L^a{ }_b $
is defined in Sect. 2.1. and $a^a$ is any real constant vector.
The operator, inducing these transformations, is then
expressed as: $ U(a,\omega) = e^{ \; i a_a p^a -\frac{i}{2}
\omega_{ab} M^{ab} }.$ The operators of the infinitesimal
translations in ordinary space $p^a$ and the Lorentz
transformations in both spaces $M^{ab}$, which are according to
Sects. 2. and 3. for spinors $M^{ab} = L^{ab} + \tilde{S}^{ab}$
and for vectors $M^{ab} = L^{ab} + S^{ab}$, close the (usual)
Poincar\' e algebra\cite{man1}:

$$ \{p^a, p^b\} = 0, \;\; \{M^{ab}, p^c\} = i(\eta^{bc} p^a -
\eta ^{ac} p^b), \eqno(4.1a) $$

\noindent
with the commutator $\{ M^{ab}, M^{cd} \} $ defined in Eq.(2.6).

Introducing the Pauli- Ljubanski vector $W^a =
\{\frac{-3}{4 \eta} \Gamma, p^a \}
$, where $\Gamma$ is defined in Eq.(2.7), one finds:

$$ W^a p_a = 0,\;\; \{p^a, W^b \} = 0,\;\; \{ M^{ab}, W^c \} =
i(\eta^{bc} W^a - \eta^{ac} W^b), \;\; \{ W^a, W^b \} = -i
\epsilon^{ab}{ }_{cd} W^c p^d, \eqno(4.1b)$$

\noindent
with $W^0 = \overrightarrow{M} . \overrightarrow{p},\;\;
\overrightarrow{W} = - \overrightarrow{M} p^0 -
\overrightarrow{K} \times \overrightarrow{p}$. The invariants of
the Poincar\' e group are then:

$$ p^2 = p^a p_a,\;\; W^2 = W^a W_a = M^2 p^2 - M_{ab} M^{cb}
p^a p_c. \eqno(4.1c)$$

\noindent
For massless representations, which we only discuss in this
paper, we have $p^2 = 0.$

\subsection{Little group and massless states.}

The representations of the Poincar\'e group
can be obtained from the irreducible representations of the
subgroup of the Poincar\' e group, called the little
group\cite{wig,fonda,han}, by the application of those generators
of the Poincar\' e group which do not form the little group.

The little group is the set of all the Lorentz transformations
which leave the four momentum of the state $k^a$ unchanged:
$k^{'a} = L^a{ }_b k^b = k^a. $ For $L^a{ }_b =
\delta^a{ }_b + \omega^a{ }_b,\;\;(\omega^a{ }_b = - \omega_b{ }^a)$,
it follows: $\; \omega^a{ }_b k^b = 0.\;\; $

If we choose $ k^a = ( k^0, 0,0,k^3 ),\;\;k^3 \ge 0,\;\;$ then
the requirement 
$\omega^a{ }_b k^b = 0 $ leaves free the three parameters of the
little group: $\omega^{1 2},\; \omega^{0 1} = \pm \omega^{1 3},
\; \omega^{0 2} = \mp \omega^{2 3},\;\;$ with the upper sign
corresponding to $k^0 = k^3 $ and the lower sign to $ k^0 =
-k^3.$   This
determines the three infinitesimal generators of the little
group: 
$l^1:= M^{01} \pm M^{13},  l^2:= M^{02} \pm M^{23},
M^{12}. $

\noindent
They fulfil the commutation relations
$ \{ l^1, l^2 \} = 0,\;\; \{ M^{1 2}, l^1 \} = il^2,\;\; \{M^{1
2}, l^2 \} = -il^1. $

The invariant of the little group is $ l^2 = (l^1)^2 + (l^2)^2,
$ while $W^a$, if applying on the little group representation,
has the form $ W^a = k^3 ( M^{1 2}, l^2, -l^1, \pm M^{1 2} ),$
where again the $+1$ is taken if $k^0 = k^3 $ and $-1$ if $ k^0 =
-k^3.$ 

Paying attention on those representations which are eigenvectors
of $l^i, i = 1,2,$ with the zero eigenvalue: $l^1 = l^2 = 0$,
only the rotations around the third axes are then nontrivially
represented and form the $U(1)$ group. In this case, since $p^2
= 0 = W^2, $ it follows:

$$ W^a = \frac{W^0}{p^0} p^a =
\frac{\overrightarrow{M}. \overrightarrow{p}}{p^0} p^a.
\eqno(4.2a)$$ 

\noindent
We easily find:

$$ \{\frac{\overrightarrow{M}. \overrightarrow{p}}{p^0}, p^a \}
= 0,\;\; 
\{ \frac{\overrightarrow{M}. \overrightarrow{p}}{p^0},
M^{ab} \} = \frac{i}{p^0} [ \eta^{a0} (W^b - \frac{W^0}{p^0}
p^b) - \eta^{b0} (W^a - \frac{W^0}{p^0} p^a )],$$
$$
\{\frac{\overrightarrow{M}. \overrightarrow{p}}{p^0}, \Gamma \}
= \frac{\overrightarrow{M.}}{p^0} [ \overrightarrow{W} -
\overrightarrow{p} \frac{W^0}{p^0}]. \eqno(4.2b) $$

\noindent
According to Eq.(4.2b) the commutators
$\{\frac{\overrightarrow{M}. \overrightarrow{p}}{p^0}, M^{ab} \}
$ and $\{\frac{\overrightarrow{M}. \overrightarrow{p}}{p^0}, \Gamma \}
$ vanish when acting in the representation space of the little
group. The operator $\frac{\overrightarrow{M}.
\overrightarrow{p}}{p^0}$ commutes with all the generators of
the Poincar\' e group and it is the invariant for the little
group representations. The invariant of the Lorentz group
$\Gamma \;( \{ \Gamma, M^{ab} \} = 0 )$ is not the invariant of
the Poincar\' e group, since $ \{
\frac{-3}{4\eta} \Gamma, p^a \} = W^a $.
However, $ \overrightarrow{L}. \overrightarrow{p} = 0 $ and
accordingly 
$\frac{\overrightarrow{M}. \overrightarrow{p}}{p^0} $ is equal to
$ \frac{\overrightarrow{\tilde{S}}. \overrightarrow{p}}{p^0}$
for spinors and $\frac{\overrightarrow{S}.
\overrightarrow{p}}{p^0}$ for vectors, depending in both cases
on only the spin and the direction of the momentum of the state.
The operator $ h:= \frac{\overrightarrow{M}.
\overrightarrow{p}}{|\overrightarrow{M} \overrightarrow{p}|} $
is called the 
helicity operator\cite{wein,fonda}.
Since $|\overrightarrow{p}| = |p^0|$ and $\{
h, \Gamma \} = 0\; $  on the little group representations, we
can choose the little group representations to be eigenstates of 
not only the helicity operator $h$ but also of the internal part
of the operator of handedness
$\Gamma $ ( in Eq.(2.7) $M^{ab} \;\;{\rm are replaced by} \;\;
\tilde{S}^{ab},\;\;{\rm or}\;\; S^{ab}$, for the
fermionic and bosonic case, respectively.) 
For the choice $k^a = ( k^0,0,0,k^3 ) $ the helicity operator 
is up to a constant equal to the operator $\tilde{S}^{1 2}$ for
spinors and to the operator $ S^{1 2}$ for vectors.

All the representations on Table I are  eigenstates of either the
operator $ \tilde{S}^{1 2} $ or the operator of handedness
$\tilde{\Gamma}$ and all the representations on Table II are
eigenstates of either the operator $ S^{1 2}$ or the operator of
handedness $\Gamma$.  To
get the representations of the little group, the states on Table
I and Table II have to be multiplied by the functions, which
are the representation states of the momentum with the
eigenvalue $k^a = ( k^0,0,0,k^3 ) $.

\subsection{Weyl-like equations.}

If the eigenvalues of the operators of handedness $\Gamma$ and 
helicity $ h = \frac{\overrightarrow{M}.
\overrightarrow{p}}{|\overrightarrow{M}| |\overrightarrow{p}|}
$ are nonzero on the 
little group representations, the eigenstates of these two operators
fulfil the Weyl-like equation((1.1)) 

$$ \overrightarrow{M}.
\overrightarrow{p} =  \xi \Gamma p^0,\eqno(4.3) $$

\noindent
with $M^{ab} = \tilde{S}^{ab}$ for spinors and $M^{ab} = S^{ab}$
for gauge  three vectors with spin $1$, with
accordingly defined $\Gamma$ and with $\xi = \frac{1}{2}$ for
spinors and 1 for vectors. While the two scalars from Table II
trivially fulfil Eq.(4.3), the four vectors from Table II, having
nonzero helicity and zero handedness,  do not at all.

We easily see that on the little group representations the
square of Eq.(4.3) leads to the equation $ p^2 = 0$ for either
the spinors or the three vectors. For spinors the square of 
Eq.(4.3) leads to the equation $ p^2 = 0 $ already on the
operator level. In addition, if multiplying Eq.(4.3) in the spinor
case by $\tilde{\Gamma}$, the Dirac equation for massless
spinors $\tilde{\gamma}^a p_a = 0 $, presented in Sect. 3, follows.

Introducing the four vector ${\cal M}^a = (\xi \Gamma,
\overrightarrow{M} ) $, again with $M^{ab} = \tilde{S}^{ab}$, 
$\xi = \frac{1}{2} $ for
spinors and $M^{ab} = S^{ab}$, $\xi = 1 $ for vectors and for
accordingly 
defined $\; \Gamma,\;$ one can write

$$ {\cal M}^a p_a = 0. \eqno(4.3a) $$

For the spinor states the operator equation (4.3a) can be
written in the matrix form, if  (for each of the four
four vectors on Table I separately) the operators
$\overrightarrow{\tilde{S}}$ and 
$\tilde{\Gamma}$ are replaced by the four by four matrices,
presented in Eq.(2.11a). Since the corresponding matrix equation
has the block diagonal form, we may further replace it by the
two Weyl equations for massless spinors:  $ \sigma^a{
}_{\pm} p_a = 0$, with $ \sigma^a{
}_{\pm} = (\pm I, 
\overrightarrow{\sigma} )$, where $ \overrightarrow{\sigma}$ are
the Pauli matrices and $ I $ is the two by two unit matrix, while
($+$) staying for the left and ($-$) for the right handed spinors,
respectively.

One can proceed equivalently also for the gauge spin 1 vector
states. The matrix form of the operator equation (4.3) follows,
if the operators are replaced by six by six matrices from Eq.(2.11b).
Since  the matrix equation has the three by three block structure,
we may further replace it by the two matrix equations: $\;\; s^a{
}_{\pm} p_a = 0,\;\;$  with $ \;\;s^a{
}_{\pm} = ( \pm I, \overrightarrow{s} )\;\; $, for the left $(+)$
and the right ($-$) handed vectors, respectively. Now 
$ I$ is the unit three by three matrix and $\overrightarrow{s}$
are the three by three matrices, which define the adjoint
representations of the group $SU(2)$ and are presented in Eq.
(2.11b). 

States, presented in Table I and Table II, are the spin part of the 
little group representations for the special case of momentum
eigenvalues $ k^a = ( 
k^0,0,0,k^3).$  For a general case  $ k^a = (
k^0,\overrightarrow{k}),\;\; $ 
superpositions of states from Table I (for spinors) and Table II
(for vectors) solve  Eq.(4.3).
 
In the case of spinors the solutions of Eq.(4.3) for $ k^a = (
k^0,\overrightarrow{k}) $, expressed by the first four vectors
from Table I ( one proceeds similarly also for the rest three
cases) $ \langle\theta| { }^{\alpha}\tilde{\varphi}_i \rangle, $
with $ \alpha = 
1 $ for left handed spinors and $ \alpha = 2 $ for right
handed spinors and $ i = 1,2, $ is as follows\footnote[8]{For all
the four cases the odd $\alpha$ corresponds to the left handed
spinor, the even $\alpha$ to the right handed spinor.}:

$$ \langle x,\theta|{ }^r \tilde{\varphi}_{\pm}\rangle = {\cal N}(x,k) 
\left\{ 
\begin{array}{clc}

 \langle\theta|{ }^{\alpha}\tilde{\varphi}_1\rangle 
+ \beta(k)
\langle\theta|{ }^{\alpha}\tilde{\varphi}_2\rangle&, &   
{\rm for}\; k^0 = r |\overrightarrow{k}|, h=1\\
-\beta^{*}(k) \langle\theta|{ }^{\alpha}\tilde{\varphi}_1\rangle
+  \langle\theta|
{ }^{\alpha}\tilde{\varphi}_2\rangle&, &   
{\rm for}\; k^0 = -r |\overrightarrow{k}|, h=-1 
\end{array}
\right\}. \eqno(4.4a)
$$

\noindent
We choose $k^3 = |k^3|$, while $\beta(k) = \frac{k^1 +
ik^2}{|k^0| + k^3}$, $\beta^{*}(k)$ means complex conjugate
value to $\beta(k)$,   $ N(x,k) = \sqrt{\frac{k^3 + 
|k^0|}{2|k^0|}} \langle x|k\rangle $ represents the normalization and
the plane wave part. Solutions fulfil the condition 
$(k^0)^2 = (\overrightarrow{k})^2 $ and correspond to left
handed spinors for
$ r = 1$ and  $\alpha = 1$
and to right handed spinors for $r = -1$ and $\alpha = 2.$ The
two helicity states have the index $+$ ( $h=1$) and $-$ ($h=-1$).

In the case of the left and right handed three vectors from
Table II, the solutions for a general case $ k^a =
(k^0,\overrightarrow{k}) $, for the left ( 
$ \alpha =
1, r = 1, $) and for the right handed ($ \alpha = 2, r = -1$ )
vectors with the helicities $ h=1$ ($+)$ and $ h=-1$ ($-$) are as
follows:

$$ \langle x,\theta|{ }^r \varphi_{\pm}\rangle = {\cal N}(x,k) 
\left\{\!\! 
\begin{array}{r}

 \langle\theta|{ }^{\alpha}\varphi_1\rangle 
+ r \sqrt{2} \beta(k) 
\langle\theta|{ }^{\alpha}\varphi_2\rangle +  (\beta(k))^2 
\langle\theta|{ }^{\alpha}\varphi_3\rangle, \;\;\;\;\;\;\;\;\\
{\rm for}\; k^0 = r |\overrightarrow{k}|, h=1\\
(\beta^{*}(k))^2 \langle\theta|{ }^{\alpha}\varphi_1\rangle
+ r \sqrt{2}\beta^{*}(k) \langle\theta|{ }^{\alpha}\varphi_2\rangle
+ \langle\theta|{ }^{\alpha}\varphi_3\rangle, \;\;\;\;\;\;\;\;\\
{\rm for}\; k^0 = -r |\overrightarrow{k}|, h=-1 
\end{array}\!\!
\right\}. \eqno(4.4b)
$$

\noindent
The functions ${\cal N}(x,k)$, $\beta(k)$ and $\beta^{*}(k)$
are the same as in the spinorial case. 
We again choose  $k^3 = |k^3|$ and assure that the condition $(k^0)^2 =
(\overrightarrow{k})^2 $ is fulfilled. Again only two (rather
than three) solutions exist for each handedness, one with
helicity $h=+1$, the other with helicity $h=-1$. The former
goes  in the
limit $\overrightarrow{k} \longrightarrow (0,0,k_3) $ to the vector
with spin $1$, the latter to the vector with spin $-1$.

The states of Eq.(4.4b) are one component wavefunctions with well
defined handedness and helicity\cite{fonda,gour,bd,han}. To
come from these one component 
objects to four component objects $A^{a}_{h r} $, which have well
defined handedness ($r$) and helicity ($h$) and which fulfil
the supplementary condition $ \;p_a A^a_{hr} = 0\;$, the one
component wave functions from Eq.(4.4b) have to be multiplied
by a part $\;e^{a}{ }_h (k)\;$ which depends on the helicity ($h$),
on momentum$(k^a)$ and not on the handedness and which fulfils
the condition $\;e^{a}_{h}(k)k_a = 
0\;$. We see that neither handedness nor helicity of states change
with the Lorentz transformations.  The orthonormalization
condition $ \;e^{*}_{ha} (k) 
e^{a}_{h^{'}}(k) = - \delta_{h h^{'}}\;$, where $(*)$ means complex
conjugation, is imposed in addition. If one looks for the space
like unit vectors $n_i,\;\; 
i=1,2, \;\;$ which both fulfil the orthonormalization equations
$ \;\; n_1{ }^a n_2{ }_a = -\delta_{1,2},\;\;$ and $ n_i{ }^a k_a =
0,\;\;$ one finds $\;\;e^{a}_{ \pm }(k) =
\frac{1}{\sqrt{2}}\;\; ( \pm n_1{ }^a - i n_2{ }^a 
), \;\;$ so that $\;\; A^{ar}_{\pm} = e^a_{\pm}(k) \langle x,\theta|{ }^r
\varphi_{\pm}\rangle$. 

One also remarks that the solution is not unique: for any
chosen set $\;\;n_i{ }^a, i=1,2 $, which solve our problem, the
infinite set of vectors $\;\;n_i{ }^a + \lambda_i  k^a,\;\; i=1,2, $
where $\lambda_i $ are any function of $ k^2$, is also a solution. This
demonstrates the gauge invariance of the  solutions $e^{a}_{h}(k)$
and correspondingly of $A^{a}_{h r}.$

\subsection{Discrete Lorentz symmetries.}

Starting from an initial vector one can find all 
vectors belonging to the same irreducible representation of the
Lorentz group by applying the generators of the homogeneous Lorentz
group. The generators of the extended Lorentz group, space and
time inversion, bring us to other representations.

According to the fact that space has in our approach (the same
number of) ordinary and Grassmann coordinates, the space and the
time inversion in ordinary and Grassmann space occur. We define
the space inversion operators for 
each of the two types of coordinates as follows:  
 
$$ I^x_s {x^a \choose p^a} (I^x_s)^{-1} = {x_a \choose
p_a}; \;\;\;  I^{\theta}_s {\theta^a \choose p^{\theta a}} (I^
\theta_s)^{-1} = {\theta_a \choose
p^{\theta}_a},\eqno(4.5a) $$ 

\noindent
The space inversion operator for the hole space is accordingly the
product of the 
above two operators: $ I_s = I^x_s I^{\theta}_s.$

We define equivalently also the time inversion operators:

$$ I^x_0 {x^a \choose p^a} (I^x_0)^{-1} = {-x_a \choose
-p_a}; \;\;\;  I^{\theta}_s {\theta^a \choose p^{\theta a}} (I^
\theta_s)^{-1} = {-\theta_a \choose
-p^{\theta}_a},\eqno(4.5b) $$ 

\noindent
The require that $p^0$, which means the energy, does not change sign
under time inversion, can be fulfilled if the time inversion operator 
is antilinear. The time inversion operator for the whole space
is therefore: $ I_0 = I^x_o I^{\theta}_0 K,$ where $K $ means
complex conjugation. Accordingly it follows: $ I_0 x^a
(I_0)^{-1} = -x_a,\;\; I_0 p^a (I_0)^{-1} = p_a,\;\;I_0 \theta^a
(I_0)^{-1} = -\theta_a,\;\;I_0 p^{\theta a}(I_0)^{-1} =
p^{\theta}_a. $ 

We see by inspection of Table I and Table II that space
inversion $ I_s $ transforms left handed spinors into right
handed ones and left handed vectors into right handed ones. Time 
inversion $ I_0 $ does not change handedness of states, either
spinorial from Table I or vectorial of Table II. In the
vectorial case we see that the time inversion transforms each of
the two three vectors into themselves.
The two scalars and the two four vectors of Table II have
handedness equal to zero.

\section{Interacting massless fermions and gauge bosons.}

We discuss in this sections those properties of the interacting
fields, which are connected  with only spin degrees of freedom.
We don't pay attention on charges. As long as the spin 
manifests as the internal degree of freedom under the
application of the Lorentz group $SO(1,3)$, and the spin and
the charge degrees of freedom are independent, states with 
different charges behave equivalently under the application of
the generators of the Lorentz transformations. Using our approach, which 
describes the internal degrees of freedom of fields as dynamics
in Grassmann space, we demonstrate that left handed spinors only
interact with left handed gauge vectors and right handed spinors
only interact with right handed gauge vectors, what is expected
also from the group theoretical point of view.    

According to what we said in Sect. 3.,  spinors always "see"
coordinates $\; \theta^a\;$ proportional to their conjugate
momenta $ 
p^{\theta a}$. Because of that in the equations of motion for
spinors all $ \theta^a $ should be replaced by $\frac{1}{2}
\tilde{a}^a$. The interacting term between
spinors and gauge vectors follows, if in the 
equation for spinors $\;\; \overrightarrow{\tilde{S}} .
\overrightarrow{p} = \frac{1}{2} \tilde{\Gamma} p^0 \;\;$
(Eq.(4.3)), 
the kinetic momentum $p_a$ is replaced by the canonical one:
$p_{0a} = p_a - A_{a}{ }_{ r} $. Being interested in only
handedness, we drop the helicity index. According to what we
have said above, $A^{a}_{r}(\theta) $ has to be replaced by $
A^{a}_{r}(\tilde{a})$, where the spin part of $
A^{a}_{r}(\theta)$ is presented in Sects. 2.7 and
4.2.  For the corresponding Lagrange term, 
describing the local interaction of a spinor and a gauge field, 
we accordingly have 
$\; \tilde{\cal L}_{int} = \langle{ }^r \tilde{\varphi}|x,\theta\rangle
\; \tilde{S}_a A^{a}_{r^{'}} \;\langle x,\theta|{
}^{r^{"}}\tilde{\varphi}\rangle,\;$ 
where  handedness $r, r^{'}, r^{"}$ can take all possible values.
Candidates for the spin part of spinor functions
$\langle\theta|{ }^r \tilde{\varphi}\rangle $ are all functions from Table
I, half of them left and half of them right handed,
and for gauge vector functions $A^a_r$ are the two three
vectors, one of the left and one of the right handedness from Table II.

In order to  evaluate the
$\;\;\tilde{\cal L}_{int} $, we first rewrite the two
three vectors in 
terms of $\tilde{a}^a. $ Then we find, if defining:$\;\;
\tilde{S}^{\pm}:= \tilde{S}^{23} \pm i\tilde{S}^{31}, \;\;
\tilde{K}^{\pm}:= \tilde{S}^{01} \pm i\tilde{S}^{02}, $  with
$\tilde{S}^{ab}$ from Eq.(2.9) and if  using the relation (Eq.2.5b)
$\tilde{a}^a \tilde{a}^b 
= -2i \tilde{S}^{ab}$, for $ a \neq b$, the followig
expressions \\
$ \langle\tilde{a}| { }^1\varphi_1\rangle = \frac{-1}{4}(\tilde{S}^+ + i\tilde{K}^+),\;\;
\langle\tilde{a}| { }^1\varphi_2\rangle = \frac{-\sqrt{2}}{4} (\tilde{S}^3 +
i\tilde{K}^3),\;\; 
  \langle\tilde{a}| { }^1\varphi_3\rangle = \frac{1}{4}(\tilde{S}^- + i\tilde{K}^-),\;\;\\
  \langle\tilde{a}| { }^2\varphi_1\rangle = \frac{1}{4}(\tilde{S}^+ - i\tilde{S}^+),\;\;
\langle\tilde{a}| { }^2\varphi_2\rangle = \frac{-\sqrt{2}}{4}(\tilde{S}^3 -i
\tilde{K}^3),\;\; 
  \langle\tilde{a}| { }^2\varphi_3\rangle = \frac{-1}{4}(\tilde{S}^- - i\tilde{K}^-).$\\
\noindent
Taking these expressions for the left (the first three) and
the right (the second three) handed three vectors into account, one
easily checks with the help of  Eq.(2.9)  that only terms
in which left handed spinors 
interact with left handed vectors and right handed spinors
interact with right handed vectors give a nonzero contributions 
to the $\tilde{\cal L}_{int}$. 
Since the spinor states from Table I are all orthonormalized
(Sects. 2.5 - 2.7.), integration over $\theta $ coordinates permits only
terms with spinors of the same handedness.

The interacting part of the Lagrange density has accordingly the
form: 

$$ \tilde{\cal L}_{int} = \tilde{\varphi}^{+}_{L}\;
\tilde{S}^a A_{a L} \tilde{\varphi}_{L}
+ \tilde{\varphi}^{+}_{R}\;
\tilde{S}^a A_{a R} \tilde{\varphi}_{R}.\eqno(5.1)$$ 

\noindent
with $\tilde{\varphi}_{L,R} = \langle x,\theta|{ }^{L,R}\tilde{\varphi}\rangle$,
$\tilde{\varphi}^{+}_{L,R} = \langle{ }^{L,R}\tilde{\varphi}|x,\theta\rangle$,
where $L,R$ stays for the left and the right handedness,
respectively. 
Eq.(5.1) can also be understood  from the group theoretical
point of view: the group $SO(1,3)$, manifesting the
$SU(2) \times SU(2)$ structure, determines  
left and right handed fundamental and left and right handed
adjoint representations. To left (right) handed fundamental
representations, the left (right) handed adjoint representations
correspond. Accordingly, only a left (right) handed gauge boson can
change spin of a left (right) handed fermion.   

One can also find that the true scalar can not influence the
spin part of spinor wave functions, while the pseudo scalar
multiplies the left handed spinors by $-1$ and the right handed
spinors by $+1$. Accordingly the two superpositions of the two
scalars from Table II 
$\;\; \frac{1}{2} ( \langle\tilde{a}|{ }^3 \varphi_1\rangle \pm
\langle\tilde{a}|{ }^4 
\varphi_1\rangle )$ multiply spinors by $ 0$ or $1$.

One can further check that the two four vectors from Table II,
belonging to the 
representations $(\frac{1}{2}, \frac{1}{2})$ and having $S^2 =
\frac{3}{2} $ (rather than $2$),  change handedness of spinors.

\section{ Discussions and conclusions. }

Following the approach in which all the internal degrees of
freedom of fields are determined within the Grassmann space, we
found the Weyl-like equations not only for massless spinors but
also for massless gauge vectors: $\overrightarrow{M} .
\overrightarrow{p} = \Gamma p^0,$ where $\overrightarrow{M}$ and
$\Gamma $ are operators which determine the spin and the
handedness for either spinors or vectors, respectively ( $M^{ab}
= \tilde{S}^{ab}$ 
for spinors and $M^{ab} = S^{ab}$ for vectors).
Accordingly the eigenstates of these two types of equations have
well defined 
not only helicity $( h = \frac{\overrightarrow{M} .
\overrightarrow{p}}{|\overrightarrow{M}| |\overrightarrow{p}|}) $
but also handedness $(\;\Gamma\;)$. Massless vector fields
demonstrate in addition the gauge invariance, as a consequence of
the fact that two rather than three superpositions of basic
states fulfil the condition $\;p^2 = 0,\;$ for each of the
two types of states. In the
representation 
with well defined helicity and handedness are the operators,
representing the generators of the Lorentz transformations and
the handedness for the gauge vector fields
the six times six 
matrices with three times three block diagonal structure ( which
demonstrates  the
left and the right handedness, respectively), while
for the spinor case they are four times four matrices with two
times two diagonal structure ( again demonstrating the left and
the right handedness, respectively). Among the vector representations
we found also the true and the pseudo scalar of the zero
handedness, trivially fulfilling the above equation.
We further found, as expected, that the space reflection
operator in Grassmann space transforms left handed fields into
right handed ones and that the time inversion 
doesn't change the handedness of fields.

Looking at the interaction terms among spinors and gauge vectors
we found  that left handed spinors only interact
with left handed gauge vectors, while right handed spinors only
interact with right handed gauge vectors. 
The sum and the difference of the true scalar and the
pseudoscalar manifest, when interacting with spinors, as
projectors. 

The experimentally measured gauge massless fields, the photon
and the gluon, having well 
defined parity rather than handedness, are described
by  the sum of states
of the left and the right handed three vectors, presented in
this paper. Such states still
have well defined helicity and still demonstrate the gauge
invariance, but are no longer solutions of the Weyl-like equations. 
On the other side, the weak boson field, when interacting with
fermions, 
only demonstrates its left handedness, while the quarks and the
leptons manifest accordingly  as left
handed weak doublets. 

Since handedness concerns the spin degrees of freedom, the
answers to the questions why we only see photons and gluons with
the positive parity and not also their partners with the
negative parity, why weak bosons interact only with left handed
fermions and why the left handed neutrino and antineutrino only
are measurable, are to our understanding all  connected.
Also the answer to the question, how the break of left-right symmetry 
occurs, should depend on properties of photons, gluons, weak
bosons and all 
fermions. Since fields of different charges demonstrate the
handedness in different ways, only the unified approach to  spins and
charges can  help to answer these questions.

In ref.\cite{pati,moha} the left-right symmetry of weak
interaction is studied and it is made manifesting by the
introduction of the two types of 
gauge $ SU(2)$ fields, one of the left and the other of the
right handedness. But handedness is the property of the
Lorentz $SO(1,3)$ group, and has nothing to do with the weak
charge, or any other charge, unless spins and charges are
unified. In these models\cite{pati,moha} the 
left-right symmetry 
is broken through the Higgs mechanism with the two types of
scalars, one of the left and one of the right handedness. But
scalars have, as we showed, zero handedness.

The approach was proposed\cite{man1,man2,man3}, in which all the
internal degrees of 
freedom, spins and charges, are described in an unique way 
within the Grassmann space. It offers the way of understanding what
is common to handedness and charges. Indeed,
in a $SO(1,14)$ multiplet, demonstrating the subgroups $SO(1,3),
SU(3), SU(2), U(1)$, the left handed $SU(2)$ doublets
together with right handed $SU(2)$ singlets appear, while the
right handed $SU(2)$ doublets appear together with the left
handed singlets\cite{ana}. This work is in progress.

We should comment at the end, what the four times two two
component Weyl spinors could represent. We think that they 
represent the four (rather than three) families of quarks and
leptons\cite{man1}. To show this a further study is needed.

\section{Acknowledgement. } This work was supported by Ministry of
Science and Technology of Slovenia.
One of the authors would like to thank very much  the
theoretical group at the University of La Plata, where the part
of this work was done, for the stimulating working atmosphere.


\begin{thebibliography}{9}

\bibitem{wig} E.P. Wigner, Ann. Math.{\bf 149}, 40 (1939),
\bibitem{sch} S.S.Schweber, An introduction to relativistic
quantum field theory, Ed. Harper and Row publisher, New York
(1962),
\bibitem{fonda} L. Fonda, G.C. Ghirardi, Theoretical physics
vol.1., Symmetry principles in quantum physics, chapter 5,
Marcel Dekker, New York, 1980,
\bibitem{bd} J.D. Bjorken, S.D. Drell, Relativistic quantum
fields, chapter 14, Inernational series in pure and applied
physics, McGraw-Hill book company 1965, New York at al,
\bibitem{gour} M. Gourdin, Basics of Lie groups, Editions
fronti\' eres, Gif sur Yvette, 1982,
\bibitem{wein} S. Weinberg, Phys. Rev. {\bf 135 b}, B1049
(1964),
\bibitem{han} D. Han, Y.S. Kim, D. Son, Phys. Rev. {\bf D 26}, 3713
(1982),
\bibitem{mirman} R. Mirman, Massless representations of the
Poincar\' e group, electromagnetism, gravitation, quantum
mechanics, geometry, Nova scienca publisher, Inc., 1995,
\bibitem{man1} N. Manko\v c Bor\v stnik, J. Math. Phys. {\bf
34}, 3731 (1993);  J. Math. Phys. {\bf
36}, 1593 (1995); N. Manko\v c Bor\v stnik, S. Fajfer, N. Cimento
{\bf B}, to appear,
\bibitem{man2} N. Manko\v c Bor\v stnik, Phys. Lett. {\bf B
292}, 25 (1992); Nuovo Cimento {\bf A 105}, 1461 (1992);
IC/91/371; Int. Jour. Mod. Phys. {\bf A 9}, 1731 (1994);
Mod. Phys. Lett. {\bf A 10}, 587 (1995); hep-th9408002;
hep-th9406083;
Proceedings of the International conference quantum systems,
New trends and methods, Minsk, 23-29 May, 1994, p. 312, Ed. by
A.O. Barut, I.D. Feranchuk, Ya.M. Shnir, L.M. Tomil'chik, World
Scientific, Singapore 1995;  Proceedings of the US-Polish
Workshop physics from Plank scale to electroweak scale, Warsaw,
21-24 Sept. 1994, p. 86, Ed. by P. Nath, T. Taylor, S. Pokorski,
World Scientific, Singapore 1995;  Proceedings of the
$7^{th}$ Adriatic meetings on High energy physics, Bri\-ju\-ni,
Croatia, 13-22 Sept.1994, p. 296, Ed. D. Klabu\v car, I. Picek,
D. Tadi\' c, World Scientific, Singapore 1995; Proceedings of
the Barut memorial conference on Group theory in physics, Tr. J.
of Phys.{\bf 21} 321 (1997),
\bibitem{man3} N. Manko\v c -Bor\v stnik, Proceedings of
the $VII^{th}$ international conference Symmetry methods in
physics, Dubna 10-18 July 1995, to appear,
\bibitem{pati} J.C. Pati, A. Salam, Phys. Rev. {\bf D 10}, 275
(1974),
\bibitem{moha} R.N. Mohapatra, J.C. Pati, Phys. Rev. {\bf D 11},
566, 2558 (1975); G, Senjanovi\` c, N. Mohapatra, Phys. Rev. {\bf D 12},
1502 (1975), R. N. Mohapatra, Unification and
supersymmetry, The frontier of quark-lepton physics,
Springer-Verlag (1986),
\bibitem{ber} F.A. Berezin and M.S. Marinov,  The methods of second
    quantization, Pure and applied physics,  Accademic press, New York,
    1966,
\bibitem{ikem} H. Ikemori,  Phys.Lett. {\bf B 199}, 239 (1987),
\bibitem{ana} N. Manko\v c Bor\v stnik, A. Bor\v stnik, in preparation.



\end{thebibliography}
\end{document}